\documentclass[pr,twocolumn,showpacs,amsmath,nobibnotes,nofootinbib,floatfix]{revtex4}
\usepackage[latin1]{inputenc}
\usepackage{color}
\usepackage{ulem}
\usepackage{graphicx}
\usepackage{bm}
\usepackage{amssymb}
\usepackage{amsmath}

\begin{document}
\title{
Spin and transport effects in quantum microcavities with polarization splitting
}

\author{M.~M.~Glazov}
\email{glazov@coherent.ioffe.ru} \affiliation{Ioffe
Physical-Technical Institute of the Russian Academy of Sciences, 194021
St.~Petersburg, Russia}
\author{L.~E.~Golub}
\affiliation{Ioffe
Physical-Technical Institute of the Russian Academy of Sciences, 194021
St.~Petersburg, Russia}

\begin{abstract}
Transport properties of exciton-polaritons in anisotropic quantum microcavities are considered theoretically. Microscopic symmetry of the structure is taken into account by allowing for both the longitudinal-transverse (TE-TM) and anisotropic splitting of polariton states. The splitting is
equivalent to an effective magnetic field acting on polariton pseudospin, and polarization conversion in microcavities is shown to be caused by an interplay of exciton-polariton spin precession and elastic scattering.
In addition, we considered the spin-dependent interference of polaritons leading to weak localization and calculated coherent backscattering intensities in different polarizations. Our findings are in a very good agreement with the recent experimental data.
\end{abstract}


\maketitle

\section{Introduction}\label{sec:intro}

Cavity polaritons are mixed states of light and matter formed as a result of the strong coupling of quantum well excitons with the photonic mode in the microcavity which embraces the quantum well. Exciton-polaritons demonstrate a wide range of spectacular phenomena caused by the combination of photonic and excitonic properties~\cite{kavokin03b}. Among those are spin effects related with an interplay of the exciton spin and photon polarization degrees of freedom~\cite{review_pol}.

The polaritonic spin states are characterized by a projection of the angular momentum on the growth axis which can be either $+1$ or $-1$. The states with a definite angular momentum  projection emit circularly polarized light, and their linear combinations correspond to the elliptically polarized light, in general. It is convenient to describe the spin dynamics of cavity polaritons in the framework of the (pseudo)spin Bloch vector whose $z$ component describes the circular polarization degree and in-plane components determine orientation of the linear polarization plane.

A driving force for polariton spin dynamics is the spin splitting of their {energy} dispersion. Acting as a wave vector dependent effective magnetic field similar to the Dresselhaus or Rashba terms in the electron effective Hamiltonian it induces the spin precession of cavity polaritons which may be directly observed {by} time-resolved photoluminescence and Faraday rotation experimental techniques~\cite{review_pol,vlad}. The powerful tool to visualize the polariton spin precession and spin splitting is the Optical spin Hall effect which consists in the linear-to-circular polarization conversion in microcavities~\cite{kavokin05a}. The angular distribution of the circular polarization degree carries information on the magnitude and the direction of an effective magnetic field acting on the polariton spin~\cite{leyder07}.

It is widely accepted that the spin splitting of the polariton states can result from the longitudinal-transverse (TE-TM) splitting of the cavity mode~\cite{review_pol,panzarini99}. This splitting is strongly wavevector dependent, and it has a symmetry of second angular harmonics because the polariton spin flip is accompanied by the angular momentum change by $\pm 2$. Another contribution to the spin splitting can be caused by the in-plane anisotropy of the microcavity which results in the splitting of the modes polarized along {two perpendicular in-plane axes}~\cite{klop,rot}. An interplay of the longitudinal-transverse and anisotropic splittings can strongly affect the spin dynamics of cavity polaritons~\cite{amo}.

Coherent effects are also very sensitive to the fine, spin-dependent structure of their energy spectrum, for review see Ref.~\onlinecite{wal_review_SST} and references therein. It was demonstrated recently that the presence of longitudinal-transverse splitting strongly affects the weak localization of exciton polaritons: the coherent backscattering can be reduced in the presence of the polariton spin splitting~\cite{polarit_prb}. So far, an analytical theory of polariton dynamics in the presence of both TE-TM and anisotropic splittings is absent.

The present paper is devoted to the theoretical study of an interplay between the longitudinal-transverse and anisotropic splittings in spin dynamics and transport properties of cavity polaritons. We apply our theory to the Optical spin Hall effect and weak localization of cavity polaritons. The analytical expressions for the polarization conversion efficiency and for the {interference-induced coherent backscattering intensities}  in microcavities are derived. The developed theory is compared with recent experimental findings~\cite{amo}.

The paper is organized as follows: in Sec.~\ref{sec:OSHE} we develop kinetic theory of Optical spin Hall effect in microcavities with allowance for the spin splitting of polariton states.  Analytical and numerical results for the polarization conversion are given. The weak localization effects are studied in Sec.~\ref{sec:WL}. The concluding remarks are presented in Sec.~\ref{sec:conclusions}.

\section{Optical Spin Hall Effect}\label{sec:OSHE}

{Experimentally detected polarization state of scattered light is described by the Stokes parameters: circular polarization degree $P_c$, and linear polarization degrees in two pairs of orthogonal axes rotated relative to each other by $45^\circ$, $P_l$ and $P_{l'}$. They are determined by the pseudospin density, $\bm S_{\bm k}$, and {particle density}, $f_{\bm k}$, of the polaritons with the in-plane wavevector $\bm k$:
\begin{equation}
	P_c(\bm k) = {S_{\bm k,z}\over f_{\bm k}}, \quad
	P_l(\bm k) = {S_{\bm k,x}\over f_{\bm k}}, \quad
	P_{l'}(\bm k) = {S_{\bm k,y}\over f_{\bm k}}.
	\label{stokes}
\end{equation}
Here  $z$ is the normal to the microcavity, and the axes $x$, $y$ lie in the microcavity plane. Hereafter we assume that the light incidence angle is small~\cite{leyder07}, therefore in calculation of Stokes parameters, Eq.~\eqref{stokes}, normal incidence can be assumed.
}

\begin{figure}[t]
\centering
\includegraphics[width=\linewidth]{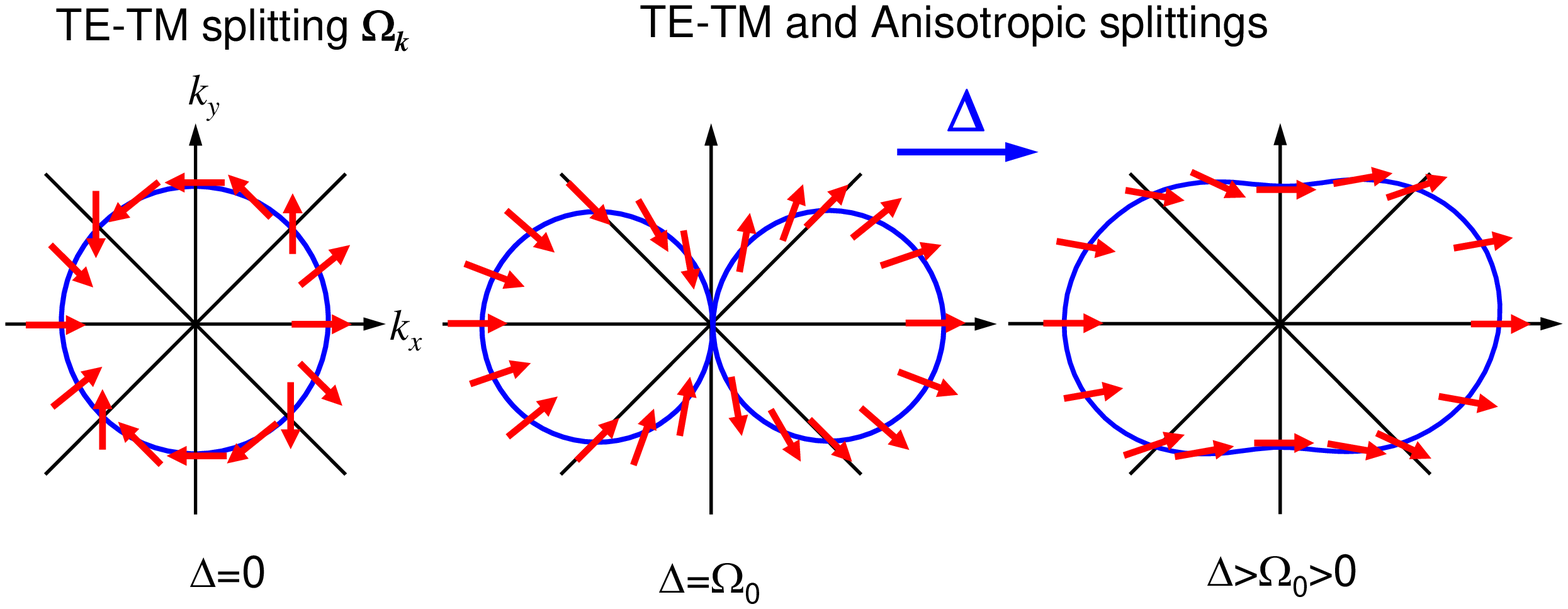}
\caption{The angular distribution of the effective magnetic
fields in $\bm k$-space at a fixed absolute value of the wave vector
$k$. Red arrows show the directions of $\bm \Omega_{\bm k}$
for different orientations of the wave vector, and blue curves
show the absolute value $\Omega_{\bm k}$.
}
\label{fig:Omegas}
\end{figure}

Classical polarization dynamics in anisotropic microcavities is described by kinetic equation for the pseudospin density of the polaritons
\begin{equation}
	\label{kin_eq}
	{{\bm S}_{\bm k} \over \tau_0} + {\bm S}_{\bm k} \times {\bm \Omega}_{\bm k}
	+ {{\bm S}_{\bm k} - \left< {\bm S}\right> \over \tau_1} = {\bm g}_{\bm k}.
	\end{equation}
Here $\tau_0$ and $\tau_1$ are the lifetime and elastic scattering times of exciton-polaritons, respectively, ${\bm g}_{\bm k}$ is the generation rate, and the angular brackets denote averaging over directions of $\bm k$. We neglect all non-linear effects caused by the polariton-polariton interaction as well as 
the inelastic scattering processes. 
The effective Larmor precession vector ${\bm \Omega}_{\bm k}$ lies in the cavity plane. It has two contributions, one with a fixed direction results from the structural anisotropy~\cite{kavokin03b,panzarini99,klop,rot}, another containing the second angular harmonics describes TE-TM splitting of the eigenmodes in ideal microcavities:
\begin{equation}
\label{Omega}
	{\bm \Omega}_{\bm k} = \bm \Delta + \Omega_0 (\cos{2\varphi},\sin{2\varphi}).
\end{equation}
Here $\varphi$ is an angle between $\bm k$ and $x$-axis, and it is assumed in what follows, that $\bm \Delta\parallel x$.
Quantities $\Omega_0$ and $\Delta$ are some functions of the wave vector absolute value $k$, which is assumed hereafter to be fixed: $k=k_0$.
The precession frequency  is anisotropic since both $\Omega_0$ and $\Delta$ are nonzero:
\begin{equation}
	\Omega_{\bm k} = \sqrt{\Omega_0^2 + \Delta^2 + 2\Omega_0\Delta\cos{2\varphi}}.
\end{equation}
The angular {dependence} of the vector ${\bm \Omega}_{\bm k}$ is plotted in Fig.~\ref{fig:Omegas} for three important cases: $\Delta=0$, $\Delta=\Omega_0$, and $\Delta > \Omega_0$.
It is worth to mention that a microcavity grown e.g. from zinc-blende lattice semiconductors possesses, in general, $C_{2\rm v}$ point symmetry group. In such a case the coefficients at $\cos{2\varphi}$ and $\sin{2\varphi}$ can be different in Eq.~\eqref{Omega}. However, this difference is related with the microscopic symmetry of the crystal lattice. {We ignore it hereafter because the main effect on the polariton pseudospin splitting is caused by the Bragg mirrors~\cite{panzarini99,klop}.}

Solution of the  kinetic equation~\eqref{kin_eq} yields the pseudospin in the form
\begin{equation}
\label{solution}
	{\bm S}_{\bm k} = {{\bm F}_{\bm k} + \tau {\bm \Omega}_{\bm k}\times{\bm F}_{\bm k} + \tau^2 {\bm \Omega}_{\bm k}({\bm \Omega}_{\bm k}\cdot {\bm F}_{\bm k}) \over 1+\Omega_{\bm k}^2\tau^2}.
\end{equation}
Here $1/\tau=1/\tau_0 + 1/\tau_1$ is a total relaxation rate, and 
\begin{equation}
\label{F}
	{\bm F}_{\bm k} = {\bm g}_{\bm k} \tau + {\tau\over\tau_1} \left< {\bm S}\right>.
\end{equation}
Equation~\eqref{solution} takes a closed form if we average it over $\varphi$ and find $ \left< {\bm S}\right>$:
\begin{equation}
	\label{S_av}
	 \left< {\bm S}\right> =  \left< I_{\bm k}{\bm F}_{\bm k}\right> +\left< {\bm J}_{\bm k} \times {\bm F}_{\bm k}\right> +  \left< \hat{\bm L}_{\bm k}{\bm F}_{\bm k}\right>.
\end{equation}
Here
\[
I_{\bm k} = {1 \over 1+\Omega^2_{\bm k}\tau^2}, \quad
{\bm J}_{\bm k} = {\tau {\bm \Omega}_{\bm k} \over 1+\Omega^2_{\bm k}\tau^2},
\]
\[
L_{{\bm k},ij} = {\tau^2 \Omega_{{\bm k},i} \Omega_{{\bm k},j} \over 1+\Omega^2_{\bm k}\tau^2}.
\]

Equations~\eqref{solution}-\eqref{S_av} describe polarization dynamics in anisotropic cavities at any excitation conditions. 
There are two important limiting cases where the spin dynamics in microcavities is most brightly pronounced: the excitation of a given state ${\bm k}_0$~\cite{leyder07}
\begin{equation}
\label{single_state}
{\bm g}_{\bm k} = {\bm g}_0 \delta_{{\bm k}{\bm k}_0},
\end{equation}
which corresponds to the standard Rayleigh scattering geometry, and the case of isotropic rate~\cite{amo}
\[
{\bm g}_{\bm k}={\bm g}.
\]

\begin{figure}[t]
\centering
\includegraphics[width=0.8\linewidth]{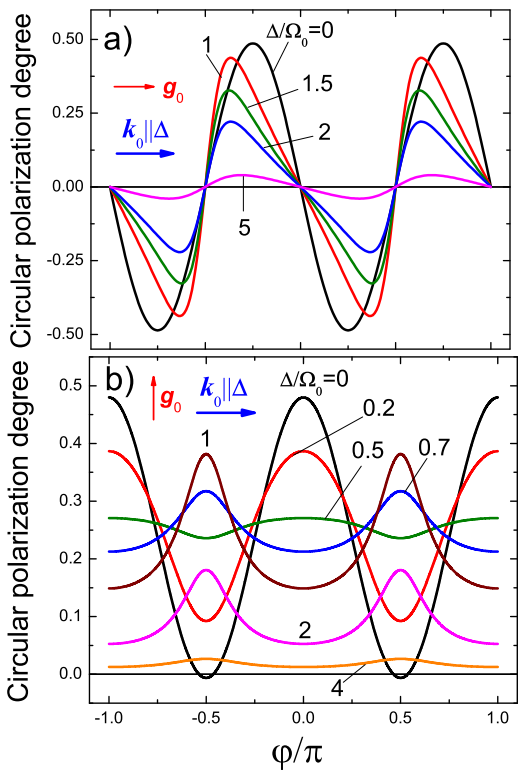}
\caption{Circular polarization degree at excitation into the states with ${\bm k}_0 \parallel {\bm \Delta}$, $\Omega_0\tau=1$, $\tau/\tau_1=0.1$.  a)~${\bm g}_0 \parallel \bm  {\bm k}_0$ and b)~${\bm g}_0 \perp {\bm k}_0$. The relative orientation of vectors $\bm g_0$ and $\bm \Delta$ is shown by arrows in the insets.}
\label{fig:OSHE}
\end{figure}

In the first case only one state on the elastic circle is excited, and the polarization in scattered states is detected. For $\Delta=0$ the problem was studied in detail in Ref.~\cite{polarit_prb}. The situation changes if the anisotropic splitting is taken into account. The angular distribution of the circular polarization degree 
given in this case by
\[
{P_c(\varphi)={S_z(\varphi)\over g_0\tau}{\tau_1\over\tau_0} }
\]
is plotted in Fig.~\ref{fig:OSHE} for excitation to the states with ${\bm k}_0 \parallel {\bm \Delta}$. At $\bm k \parallel \bm k_0$ (i.e. at $\varphi=0$) we disregard the contribution of the pump. Panel~(a) corresponds to ${\bm g}_0 \parallel {\bm k}_0$; in this case the eigenstates are excited: ${\bm g}_0 \parallel {\bm \Omega}_{{\bm k}_0}$. Panel~(b) describes the case ${\bm g}_0 \perp {\bm k}_0$, when ${\bm g}_0$ is perpendicular to ${\bm \Omega}_{{\bm k}_0}$, cf.~Fig.~\ref{fig:Omegas}.

Figure~\ref{fig:OSHE}(a) shows that the circular polarization degree has two maxima and two minima whose amplitudes decrease with an increase of the anisotropic splitting $\Delta$. At small $\Delta\ll \Omega_0$ these extrema are positioned at $\varphi= \pm \pi/4$, $\pm 3\pi/4$ corresponding to the scattering angles where $\bm \Omega_{\bm k}$ and $\bm g_0$ are orthogonal which leads to the highest conversion efficiency~\cite{kavokin05a}. With an increase of $\Delta$ the conversion efficiency is reduced because the overall spin splitting tends to be parallel to $\bm \Delta \parallel \bm g_0$. Indeed, if $\Delta \gg \Omega_0$ the conversion is caused by the TE-TM splitting solely, but the circular polarization degree is strongly suppressed due to the fast precession of the pseudospin in the plane perpendicular to $\bm \Delta$ similarly to the Hanle effect.

The situation drastically changes if one excites the cavity with {$\bm g_0 \perp {\bm k}_0$ (${\bm k}_0 \parallel \bm \Delta$)}, Fig.~\ref{fig:OSHE}(b). In such a case the initial state is not an eigenstate of the system even if $\Delta=0$, {cf. Fig.~\ref{fig:Omegas}}. Hence, the non-zero angular averaged circular polarization $\left< P_c \right>$ appears, and the conversion efficiency is reduced with an increase of the anisotropic splitting $\Delta$ due to the faster pseudospin precession. The minima of the conversion efficiency positioned at $\varphi=\pm \pi/2$ for $\Delta=0$ are converted into maxima with an increase of $\Delta$. This happens because the total fields $\bm \Omega_{\bm k}$ at $\varphi=0$ and at $\varphi=\pm \pi/2$ are opposite for $\Delta=0$, while for $\Delta \gg \Omega_0$ they are equal.

Now we turn to the isotropic generation. First, it is instructive to analyze  the angular-integrated degree of emission polarization. 
In the case of $\bm g \parallel \bm \Delta$ the angular averaged circular $\langle P_c\rangle$ and linear  $\langle P_{l'}\rangle$ polarizations 
vanish from the symmetry arguments. The relaxation of the parallel to $\bm \Delta$ pseudospin component {$S_x$} is suppressed by the presence of the anisotropic splitting similarly to the suppression of the D'yakonov-Perel' spin relaxation by the Larmor effect of the magnetic field. Hence, {$S_x$}  increases with the increase of $\Delta$, and the linear polarization degree $\langle P_l\rangle$ reaches $1$ at $\Delta\tau\gg 1$, $\Delta\gg \Omega_0$.

\begin{figure}[t]
\centering
\includegraphics[width=0.8\linewidth]{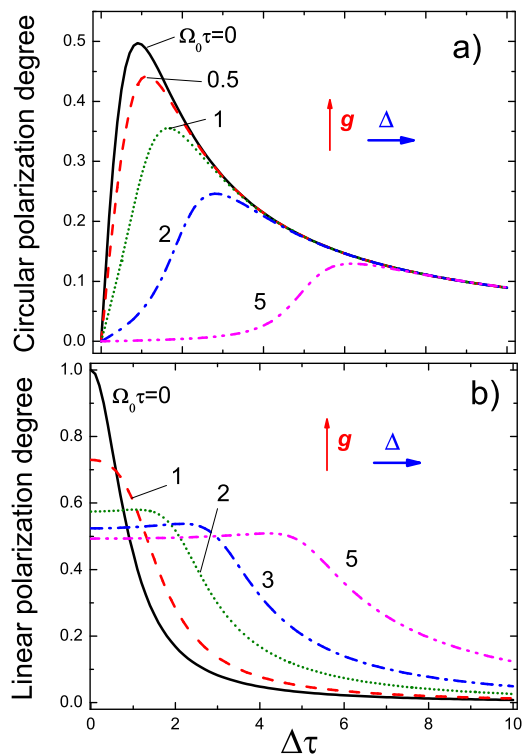}
\caption{Dependences of circular polarization degree $\left< P_c \right>$ (a) and linear polarization degree 
$\left< P_{l'} \right>$ (b) on $\Delta\tau$ at isotropic excitation with ${\bm g}\perp {\bm \Delta}$,
$\tau/\tau_1=0.1$.}
\label{fig:Polarizations}
\end{figure}

Figure~\ref{fig:Polarizations} represents the analysis for 
orientation of the generation vector ${\bm g}$ 
perpendicular to ${\bm \Delta}$. 
Panel~(a) presents 
the circular polarization degree, $\left< P_c \right>$, and
panel~(b) shows the 
linear polarization degree $\left< P_{l'} \right>$ [cf. Eq.~\eqref{stokes}]:
\[
\left< P_c \right>={\left< S_z \right>\over g\tau_0}, \qquad \left< P_{l'} \right>={\left< S_y \right>\over g\tau_0}. 
\]
It can be seen from Fig.~\ref{fig:Polarizations}(a) that the angular-integrated circular polarization degree first increases with an increase of $\Delta$. This happens because the anisotropic splitting acts as a constant magnetic field and induces the conversion of {perpendicular to $\bm \Delta$ in plane} pseudospin component to the {out of plane} component. Further increase of $\Delta$ results in suppression of the circular polarization degree due to the spin precession, similarly to the results shown in Fig.~\ref{fig:OSHE}(a). Accordingly, {the in plane} pseudospin component is decreased by the effective magnetic field $\bm \Delta$ in agreement with Fig.~\ref{fig:Polarizations}(b).

Then, we analyse the angular distribution of the circular polarization degree 
\[
P_c(\varphi)={S_z(\varphi)\over g\tau_0}.
\]
In the case $\bm g \parallel \bm \Delta$ it has the same form as for the generation to a single state {${\bm k}_0 \parallel {\bm \Delta}$} shown in Fig.~\ref{fig:OSHE}a. Indeed, as it follows from the symmetry of the problem, the angular averaged pseudospin vector $\langle \bm S \rangle$ is parallel to $\bm \Delta$, and the solutions of Eq.~\eqref{kin_eq} for $\bm g_{\bm k} \propto \delta_{\bm k \bm k_0}$ and $\bm g_{\bm k} = const$ are different by a constant factor only.
\begin{figure}
\centering
\includegraphics[width=0.8\linewidth]{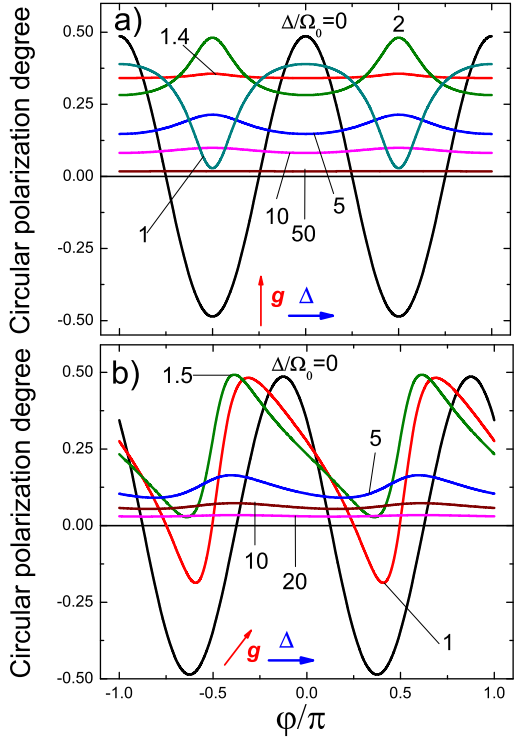}
\caption{{Circular polarization degree at $\Delta\tau=1$, $\tau/\tau_1=0.1$. a)~${\bm g} \perp {\bm \Delta}$, b)~${\bm g}$ at $45^\circ$ to ${\bm \Delta}$.} }
\label{fig:Noneigenstate}
\end{figure}
Therefore in Fig.~\ref{fig:Noneigenstate} we demonstrate the angular distribution 
$P_c(\varphi)$ for two specific orientations of the generation vector, ${\bm g} \perp {\bm \Delta}$ (Fig.~\ref{fig:Noneigenstate}a) and ${\bm g}$ at $45^\circ$ to ${\bm \Delta}$  (Fig.~\ref{fig:Noneigenstate}b).
 The main contribution to the angular dependence is given by zeroth and second harmonics, cf.~Eq.~\eqref{Omega}. 
With an increase of $\Delta$ the zeroth harmonics contribution first increases and then decreases in agreement with Fig.~\ref{fig:Polarizations}. For $\Delta \gg \Omega_0$ the angular distribution of the circular polarization degree is almost constant because the spin precession vector points along the same axis, {cf.~Fig.~\ref{fig:Omegas}}. Note, that in the case of $\bm g$ oriented by $45^\circ$ to $\bm \Delta$ the angular distribution is asymmetric with respect to $\varphi\to -\varphi$, and the asymmetry is most pronounced for comparable $\Delta$ and $\Omega_0$.

\begin{figure}
\centering
\includegraphics[width=0.8\linewidth]{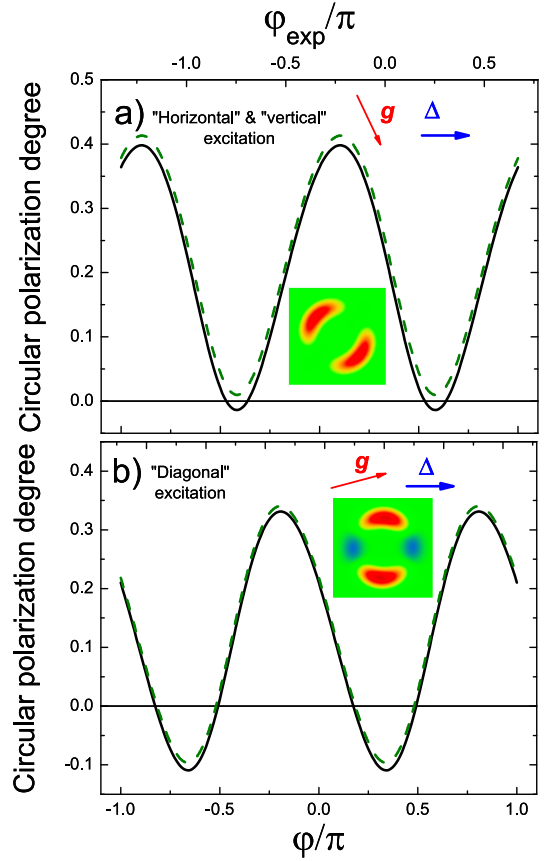}
\caption{Angular distribution of the circular polarization degree at conditions of experiment~\cite{amo}: $\Delta\tau=\Omega_0\tau=0.256$, $\tau_1\gg\tau_0$. Solid and dashed curves are calculated at $\tau/\tau_1=\tau_0/\tau_1$ equal to 0.1 and 0.2, respectively. Panel~(a) corresponds to the panels~(a,b) and (c,d), panel~(b) corresponds to the panels~(e,f) from Fig.~5 of Ref.~\cite{amo}.  Insets show color plots of the $\bm k$-space distribution of the circular polarization degree. Axes in the insets coincide with those in Ref.~\cite{amo}. }
\label{fig:Exp_compar}
\end{figure}

In order to compare our theory with Ref.~\cite{amo} we present results of calculations for their experimental conditions: we take the polariton lifetime $\tau_0 = 4$~ps, equal values of the TE-TM and anisotropic splittings, $\hbar\Omega_0 = \hbar \Delta =0.04$~meV, and the momentum scattering time $\tau_1$ 
much longer than $\tau_0$. We also take into account that {at experimental conditions and for the sample studied in Ref.~\cite{amo} the excitation was performed at $\bm g$ oriented at angles $-60^\circ$, $120^\circ$, and $30^\circ$ with respect to the vector $\bm \Delta$.} The corresponding angular distributions of the circular polarization degree are plotted in Fig.~\ref{fig:Exp_compar}, 
where
$\varphi=0$ corresponds to the direction of the vector $\bm \Delta$. Note that in Fig.~5 of Ref.~\cite{amo} the circular polarization degree is plotted vs $\varphi_{\rm exp}=\varphi-120^\circ$ (top axis in our Fig.~\ref{fig:Exp_compar}).
It is seen, that the circular polarization is almost insensitive to the value of the elastic scattering time $\tau_1$. This happens because the single scattering regime at $\tau_1\gg \tau_0$ is realized. 

One can see from Fig.~\ref{fig:Exp_compar} that the agreement between our kinetic theory and the experimental data is quite good: The circular polarization degree varies in a range of $-0.4\ldots 0.4$ 
as in Ref.~\cite{amo}. Then, as it follows immediately from the linearity of the kinetic equation~(\ref{kin_eq}), the change of polarization from ``horizontal'' to ``vertical'' (i.e. change $\bm g \to -\bm g$) results in the change of circular polarization sign in agreement with panels (a) and (c) in the experimental figure~5, Ref.~\cite{amo}. Hence we have plotted in Fig.~\ref{fig:Exp_compar}(a) only the curves for one orientation of $\bm g$ (``horizontal''). 
At ``diagonal'' excitation, Fig.~\ref{fig:Exp_compar}{b}, the averaged circular polarization degree $\left< P_c \right>$ is much smaller as compared with the panel~(a). The angular positions of the polarization maxima and minima shift to higher angles in a good agreement with the experiment. The color plots of the circular polarization degree in the $\bm k$-space shown as insets agree well with the experimental data presented in Fig.~5 of Ref.~\cite{amo}.

\section{Weak Localization Effects}\label{sec:WL}

The classical kinetic theory presented in the previous Section describes well the available experimental data on the Optical spin Hall effect in microcavities. The polaritons, however, are known to keep their coherence while propagating over large distances~\cite{kavokin03b,leyder07}. As a result, {interference} effects can come into play. The most important of those are the coherent 
phenomena leading to weak localization of polaritons~\cite{polarit_prb,gurioli:183901,liew}.

In the Rayleigh scattering experiments the angular distribution of the scattered polaritons is observed. In what follows we concentrate on the case of the single state excitation, Eq.~\eqref{single_state}. {We also assume the multiple-scattering regime: $\tau_0 \gg \tau_1 \approx \tau$.} The interference of polaritons induces the corrections to the particle number density ($\delta f_{\bm k}$) and spin density ($\delta {\bm S}_{\bm k}$):
\begin{equation}
	\label{kin_eq_WL_f}
	\delta f_{\bm k} = \sum_{\bm k'} A_0(\bm k, \bm k') f_{\bm k'},
\qquad	
\delta {\bm S}_{\bm k} = \sum_{\bm k'} \hat{A}(\bm k, \bm k') {\bm S}_{\bm k'} ,
\end{equation}
\begin{equation}
\label{Cooperon}
	A_0 = {1\over 2} \sum_{\alpha\beta} {\cal C}^{\alpha\beta}_{\beta\alpha}(\bm k+\bm k'),
	\quad	
	A_{ij} = {1\over 2} \sum_{\alpha\beta\gamma\delta} \sigma^i_{\gamma\beta}{\cal C}^{\alpha\beta}_{\gamma\delta}(\bm k+\bm k') \sigma^j_{\alpha\delta},
\end{equation}
where the Cooperon $\cal C$ is a sum of all ``fan'' diagrams, i.e. diagrams with maximally crossed scattering lines, see Ref.~\cite{polarit_prb} for details. In Eqs.~\eqref{Cooperon} we assumed that $\Omega\tau\ll 1$.

The main effect of the interference is the modification of the backscattering. Indeed, the substitution of any smooth part of $f_{\bm k'}$, ${\bm S}_{\bm k'}$ into the r.h.s of Eq.~\eqref{kin_eq_WL_f} leads to small corrections of order $1/(kl_0)^2\ll 1$, where $k$ is the characteristic polariton wavevector, and 
\[
l_0 = v\sqrt{\tau\tau_0\over 2}
\]
is the dephasing length of the polariton, coinciding with the typical displacement during the life-time $\tau_0$ for pure elastic scattering considered here.
Therefore the significant corrections are obtained only after substitution of the singular terms appeared in the distribution at the point of generation which, in the limit of $\Omega_0\tau\ll 1$, $\Delta\tau \ll 1$ read:
\[
f_{\bm k'} = g_0\tau\delta_{\bm k' \bm k_0}, 
\qquad
{\bm S}_{\bm k'} = \bm g_0 \tau\delta_{\bm k' \bm k_0}.
\]
As a result we get
\begin{equation}
	\label{WL_fin}
	\delta f_{\bm k} =   A_0(\bm k+\bm k_0) \, g_0\tau ,
\qquad	
\delta {\bm S}_{\bm k} = \hat{A}(\bm k+\bm k_0) \, {\bm g}_0\tau .
\end{equation}

It is instructive to relate the functions $A_0(q)$, $\hat A(q)$ with the spin-dependent return probabilities which describe the coherent backscattering corrections to the kinetic equation. Indeed, if one is not interested in the details of the distribution function in the wave vector scale of $1/l$ (and, hence, in the scale of $1/l_0 \ll 1/l$), one can represent the kinetic equations for the particle and spin densities as follows~\cite{polarit_prb}:
\begin{equation}
	\label{kin_eq_WL_f1}
	{f_{\bm k} \over \tau_0} 
	+ {f_{\bm k} - \left< f\right> \over \tau_1} - W_0 \left(f_{-\bm k} - \left< f\right> \right) = g_{\bm k},
\end{equation}
\begin{equation}
	\label{kin_eq_WL1}
	{{\bm S}_{\bm k} \over \tau_0} + {\bm S}_{\bm k} \times {\bm \Omega}_{\bm k}
	+ {{\bm S}_{\bm k} - \left< {\bm S}\right> \over \tau_1} - \hat{W}\left({\bm S}_{-\bm k} - \left< {\bm S}\right>\right) = {\bm g}_{\bm k},
\end{equation}
where the values of {the spin-dependent return probabilities} $W_0$, $\hat W$ are related with the functions $A_0$, $\hat A$ {as}:
\begin{equation}
	W_0 = {2l\tau_0\over k_0 \tau^2}\sum_{\bm q} A_0(\bm q),
	\qquad
	W_{ij}= {2l\tau_0\over k_0 \tau^2}\sum_{\bm q} A_{ij}(\bm q).
\end{equation}

The anisotropic microcavity has $D_{\rm 2h}$ point symmetry (or $C_{\rm 2v}$ if the microscopic structure of the crystalline lattice is taken into account) with the $C_2$ axis coinciding with the normal $z$ direction. It means that the only non-zero components of $\hat A$ are
\[
 A_{xx}, \quad A_{yy}, \quad A_{zz}, \quad A_{yz}=-A_{zy}.
\]
Note that the relation between $A_{yz}$ and $A_{zy}$ components is identical to that for the off-diagonal components of the conductivity tensor in a magnetic field $\bm B \parallel x$ (anisotropic in $\bm q$ contributions to $A_0$, $\hat A$ have extra smallness caused by the spin splitting and are neglected). 
Calculation shows that
\begin{subequations}
\label{AAA}
\begin{equation}
A_0(\bm q)= {1\over2} \left(C_0 + C_1 + {C_- - C_+\over R} \right),
\end{equation}
\begin{equation}
	A_{xx}(\bm q)= {1\over2} \left(C_0 + C_1 - {C_- - C_+\over R} \right),
\end{equation}
\begin{equation}
	A_{yy}(\bm q)=  {1\over2} \left(C_0 - C_1 + C_- + C_+ \right),
\end{equation}
\begin{equation}
	A_{zz}(\bm q)= {1\over2} \left(-C_0 + C_1 + C_- + C_+ \right),
\end{equation}
\begin{equation}
	A_{yz}(\bm q)=  -A_{zy}(\bm q)= {\Delta\tau_s\over R} (C_- - C_+).
\end{equation}
\end{subequations}
Here $R=\sqrt{1-(2\Delta\tau_s)^2}$,
\begin{equation}
\nonumber
	C_0={1\over 1 + (ql_0)^2},
	\qquad
C_1={1\over 1 + (ql_0)^2 + \tau_0/\tau_s},
\end{equation}
\begin{equation}
\nonumber
	C_\pm = {1\over 1 + (ql_0)^2 + \Delta^2\tau\tau_0 + (3\pm R) \tau_0/(2\tau_s)},
\end{equation}
and
\begin{equation}
\nonumber
	{1\over\tau_s} = {\Omega_0^2\tau\over 2},
\end{equation}
is the relaxation rate for the in-plane pseudospin components~\cite{polarit_prb}.

One can see from Eqs.~\eqref{AAA} that the values $A_0(\bm k + \bm k_0)$, $\hat A(\bm k + \bm k_0)$ which determine the angular distribution of the particles have sharp peaks at $\bm k \approx - \bm k_0$ which correspond to the coherent backscattering. The processes of the coherent scattering by an arbitrary angle are disregarded here since they contribute to the smooth part of the distribution function at $|\bm k+ \bm k_0|l, |\bm k + \bm k_0|l_0 \gtrsim 1$.

In the limit $\Delta\tau_s\ll 1$, $\Delta\sqrt{\tau\tau_0}\ll 1$, we get
\begin{widetext}
\begin{subequations}
\label{As}
\begin{equation}
\label{A0s}
	A_0(\bm q)= {1\over2} \left[
	{1\over (ql_0)^2 + 1} 
	+{2\over (ql_0)^2 + \tau_0/T_{s\perp}}
	-	{1\over (ql_0)^2 + \tau_0/T_{s\parallel}}
	\right],
\end{equation}
\begin{equation}
	A_{xx}(\bm q)= A_{yy}(\bm q)= {1\over2} \left[
	{1\over (ql_0)^2 + 1} 
	+	{1\over (ql_0)^2 + \tau_0/T_{s\parallel}}
	\right] \equiv A_{\perp}(q),
\end{equation}
\begin{equation}
\label{Azzs}
	A_{zz}(\bm q)= {1\over2} \left[
	{2\over (ql_0)^2 + \tau_0/T_{s\perp}}
	+	{1\over (ql_0)^2 + \tau_0/T_{s\parallel}}
	- {1\over (ql_0)^2 + 1}
	\right] \equiv A_{\parallel}(q),
\end{equation}
\begin{equation}
	A_{yz}(\bm q)= \Delta\tau_s \left[
	{1\over (ql_0)^2 + \tau_0/T_{s\perp}}
	-	{1\over (ql_0)^2 + \tau_0/T_{s\parallel}}
	\right],
\end{equation}
\end{subequations}
\end{widetext}
where the lifetimes are introduced for spin components parallel
and perpendicular to the growth axis $z$~\cite{polarit_prb}:
\begin{equation}
	{1\over T_{s\parallel}} = {1\over\tau_0} + {2\over\tau_s},
	\qquad
	{1\over T_{s\perp}} = {1\over\tau_0} + {1\over\tau_s}.
\end{equation}

In the absence of both the longitudinal-transverse and the anisotropic splittings the polariton pseudospin is not affected in the course of the propagation. In this case, $A_0=A_{ii}= 1/[1+(ql_0)^2]$, and, in agreement with Eq.~\eqref{WL_fin} the total number of the backscattered (at $\bm k=-\bm k_0$, i.e. $q=0$) particles is twice higher than the number of the particles scattered by an arbitrary angle. The same applies for all pseudospin components: the emission intensity in a given polarization is twice higher for $\bm k=-\bm k_0$ as compared with the intensity for the arbitrary scattering angle.

The presence of the longitudinal-transverse and anisotropic splittings qualitatively changes the situation. Although the pseudospin splittings of polariton energy spectrum do not affect their propagation as long as the splittings are much smaller as the characteristic particle energy, the interference of the particles is strongly affected. 
\begin{figure*}
\centering
\includegraphics[width=\linewidth]{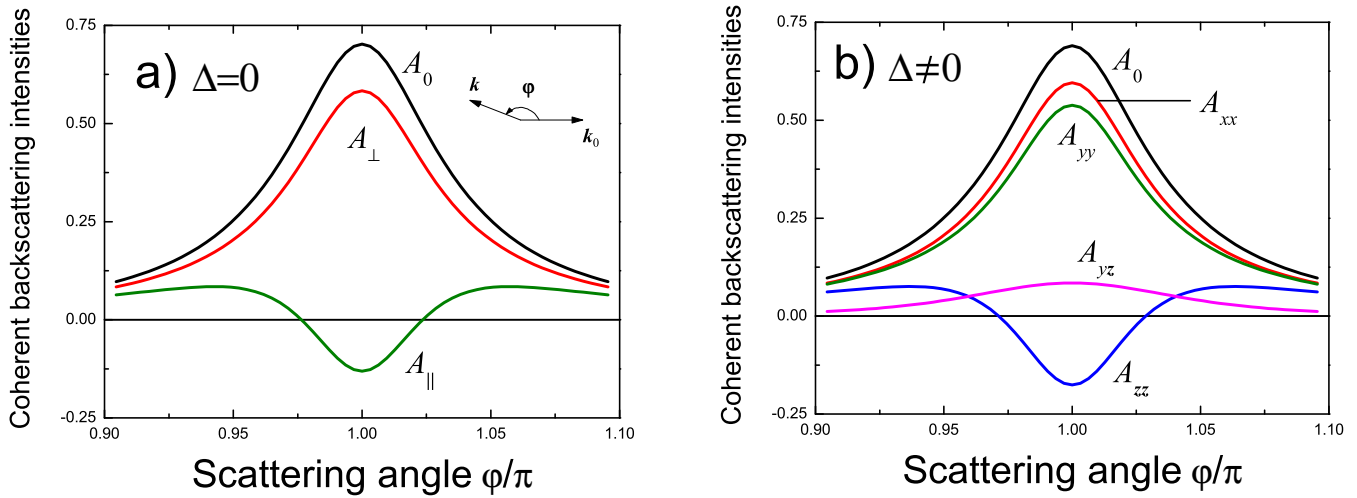}
\caption{Angular distribution of relative coherent backscattering intensities, $A_0$, $A_{ij}$ for a)~isotropic spin-splitting and b)~anisotropic spin splitting $\Delta\tau=0.02$. The parameters of the calculation are $\tau_0/\tau=100$, $kl_0=10$, $\Omega_0\tau=0.22$. }
\label{fig:WAL}
\end{figure*}
Figure~\ref{fig:WAL} shows the coherent backscattering intensities, $A_0$, ${A}_{ij}$ as functions of the scattering angle $\varphi$ calculated by Eqs.~\eqref{AAA} with $q=2k|\cos{(\varphi/2)}|$.
Firstly, we consider the case where the anisotropic splitting is absent, $\Delta=0$. It is demonstrated in Fig.~\ref{fig:WAL}a that the presence of the longitudinal-transverse splitting, $\Omega_0$, partially suppresses the interference, and the backscattering peak becomes lower and wider than at $\Omega_0=0$. In agreement with Eq.~\eqref{A0s} it consists of three contributions corresponding to different spin states of the interfering particles. Qualitative behavior of the backscattering peak in linear polarization, $A_{xx}$, $A_{yy}$ is similar. Interestingly, the backscattering peak in the circular polarization [Eq.~\eqref{Azzs} at $\Omega_0=0$] is transformed into a dip provided the longitudinal-transverse splitting is relatively strong, Fig.~\ref{fig:WAL}a. Besides, in agreement with Eqs.~\eqref{As}, only $A_{zz}$ can change its sign for relatively strong longitudinal-transverse splitting where $\tau_s\ll \tau_0$. This effect is a consequence of the fact that the real spin of exciton-polaritons is integer (the Berry phase is $2\pi$), and the anti-localization behavior is manifested in pseudospin $z$ component, unlike the case of electrons where the correction to the diffusion constant (i.e. $A_0$) changes its sign as a function of the spin splitting~\cite{polarit_prb}.

The distribution of the backscattered particles becomes even more rich if the anisotropic splitting is taken into account, $\Delta\ne 0$, see Fig.~\ref{fig:WAL}b. Clearly, if the isotropic splitting is absent, one can quantize the polariton pseudospin onto the axis $\bm \Delta$, and the interference corrections for the particle number, $A_0$, are exactly the same as in the absence of the anisotropic splitting, in agreement with Eqs.~\eqref{AAA}~\cite{wal_review_SST}. If both $\Omega_0\ne 0$ and $\Delta\ne 0$ the conversion between $y$ and $z$ pseudospin components (i.e. between the circular and diagonal linear polarizations) {described by the odd in $\Delta$ components $A_{yz}=-A_{yz}$} appears in the backscattering. Besides, as shown in Fig.~\ref{fig:WAL}b the backscattering becomes different in linear polarizations: $A_{xx}> A_{yy}$. It is a result of the fact that the dynamics of the parallel and perpendicular to $\bm \Delta$ components of the pseudospin is different. Indeed, for relatively strong anisotropic splitting, $\Delta\tau_s \gg 1$, $\Delta\tau\gg 1$ (not shown) the interference of $y$ pseudospin components (in the diagonal linear polarization) and of $z$ pseudospin components (in the circular polarization) should be absent since these components are rapidly lost as a result of the spin precession in the field $\bm \Delta$. At the same time the interference of $x$ pseudospin components (linear polarization in $xy$ axis) as well as the interference of unpolarized particles remains the same as in the absence of both longitudinal-transverse and anisotropic splittings because the eigenstates of the system correspond to the definite $x$ pseudospin projections.

\section{Conclusions}\label{sec:conclusions}

In conclusion, we have studied in detail the exciton-polariton spin dynamics with allowance for both the longitudinal-transverse splitting and the anisotropic splitting which coexist in real structures. The presence of the anisotropic splitting changes dramatically the polarization conversion in microcavities as compared with ideal isotropic systems where only TE-TM splitting is of importance. It turns out that the angular-integrated emission of the microcavity excited by linearly polarized light becomes, in general, elliptically polarized. The efficiency of the linear to circular polarization conversion depends strongly on the relation between the TE-TM splitting, the anisotropic splitting and the polariton radiative and scattering rates.

We have analyzed the effects of anisotropic splitting on the interference of polaritons caused by the weak localization/antilocalization phenomena. The spin-dependent backscattering intensities are shown to be strongly sensitive to the anisotropic splitting of polariton states. For instance, weak localization itself leads to the conversion from linear to the circular polarization in the course of polariton diffusion. 

Application of  our model to recent experimental data on Optical spin Hall effect in microcavities~\cite{amo} showed a very good agreement with the experiment.

\acknowledgments{{We thank A.V. Kavokin for the remarks on the manuscript.} This work was partially supported by RFBR, ``Dynasty'' Foundation---ICFPM  and President grant for young scientists.}


\begin{thebibliography}{24}
\expandafter\ifx\csname natexlab\endcsname\relax\def\natexlab#1{#1}\fi
\expandafter\ifx\csname bibnamefont\endcsname\relax
  \def\bibnamefont#1{#1}\fi
\expandafter\ifx\csname bibfnamefont\endcsname\relax
  \def\bibfnamefont#1{#1}\fi
\expandafter\ifx\csname citenamefont\endcsname\relax
  \def\citenamefont#1{#1}\fi
\expandafter\ifx\csname url\endcsname\relax
  \def\url#1{\texttt{#1}}\fi
\expandafter\ifx\csname urlprefix\endcsname\relax\def\urlprefix{URL }\fi
\providecommand{\bibinfo}[2]{#2}
\providecommand{\eprint}[2][]{\url{#2}}

\bibitem[{\citenamefont{Kavokin and Malpuech}(2003)}]{kavokin03b}
  A. Kavokin, J. Baumberg, G. Malpuech, F. Laussy, \textit{Microcavities}, Clarendon Press Oxford (2006).

\bibitem{review_pol} See for review I.A. Shelykh, A.V. Kavokin, Yuri G Rubo, T.C.H. Liew, and G. Malpuech, Semicond. Sci. Technol. {\bf 25}, 013001 (2010) and references therein.

\bibitem{vlad} A. Brunetti, M. Vladimirova, D. Scalbert, M. Nawrocki, A. V. Kavokin, I. A. Shelykh, and J. Bloch, Phys. Rev. B {\bf 74}, 241101 (2006).

\bibitem[{\citenamefont{Kavokin et~al.}(2005)\citenamefont{Kavokin, Malpuech,
  and Glazov}}]{kavokin05a}
\bibinfo{author}{\bibfnamefont{A.}~\bibnamefont{Kavokin}},
  \bibinfo{author}{\bibfnamefont{G.}~\bibnamefont{Malpuech}}, \bibnamefont{and}
  \bibinfo{author}{\bibfnamefont{M.}~\bibnamefont{Glazov}},
  \bibinfo{journal}{Phys. Rev. Lett.} \textbf{\bibinfo{volume}{95}},
  \bibinfo{eid}{136601} (\bibinfo{year}{2005}).

\bibitem[{\citenamefont{Leyder et~al.}(2007)\citenamefont{Leyder, Romanelli,
  Karr, Giacobino, Liew, Glazov, Kavokin, Malpuech, and Bramati}}]{leyder07}
\bibinfo{author}{\bibfnamefont{C.}~\bibnamefont{Leyder}},
  \bibinfo{author}{\bibfnamefont{M.}~\bibnamefont{Romanelli}},
  \bibinfo{author}{\bibfnamefont{J.~P.} \bibnamefont{Karr}},
  \bibinfo{author}{\bibfnamefont{E.}~\bibnamefont{Giacobino}},
  \bibinfo{author}{\bibfnamefont{T.~C.~H.} \bibnamefont{Liew}},
  \bibinfo{author}{\bibfnamefont{M.~M.} \bibnamefont{Glazov}},
  \bibinfo{author}{\bibfnamefont{A.~V.} \bibnamefont{Kavokin}},
  \bibinfo{author}{\bibfnamefont{G.}~\bibnamefont{Malpuech}}, \bibnamefont{and}
  \bibinfo{author}{\bibfnamefont{A.}~\bibnamefont{Bramati}},
  \bibinfo{journal}{Nature Physics} \textbf{\bibinfo{volume}{3}},
  \bibinfo{pages}{628} (\bibinfo{year}{2007}).

\bibitem[{\citenamefont{Panzarini et~al.}(1999)\citenamefont{Panzarini,
  Andreani, Armitage, Baxter, Skolnick, Astratov, Roberts, Kavokin,
  Vladimirova, and M.A.Kaliteevski}}]{panzarini99}
\bibinfo{author}{\bibfnamefont{G.}~\bibnamefont{Panzarini}},
  \bibinfo{author}{\bibfnamefont{L.~C.} \bibnamefont{Andreani}},
  \bibinfo{author}{\bibfnamefont{A.}~\bibnamefont{Armitage}},
  \bibinfo{author}{\bibfnamefont{D.}~\bibnamefont{Baxter}},
  \bibinfo{author}{\bibfnamefont{M.~S.} \bibnamefont{Skolnick}},
  \bibinfo{author}{\bibfnamefont{V.~N.} \bibnamefont{Astratov}},
  \bibinfo{author}{\bibfnamefont{J.~S.} \bibnamefont{Roberts}},
  \bibinfo{author}{\bibfnamefont{A.~V.} \bibnamefont{Kavokin}},
  \bibinfo{author}{\bibfnamefont{M.~R.} \bibnamefont{Vladimirova}},
  \bibnamefont{and} \bibinfo{author}{\bibnamefont{M.A. Kaliteevski}},
  \bibinfo{journal}{Phys. Solid State} \textbf{\bibinfo{volume}{41}},
  \bibinfo{pages}{1223} (\bibinfo{year}{1999}).



\bibitem{klop} L. Klopotowski, M. D. Martin, A. Amo, L. Vina, I.A. Shelykh,
M.M. Glazov, G. Malpuech, A.V. Kavokin, R. Andre, Solid State Commun. {\bf 139}, 511
(2006).

\bibitem{rot} D.N. Krizhanovskii, D. Sanvitto, I.A. Shelykh, M.M. Glazov, G. Malpuech, D.D. Solnyshkov, A. Kavokin, S. Ceccarelli, M. S. Skolnick, and J. S. Roberts, Phys. Rev. B {\bf 73}, 073303 (2006).

\bibitem{amo} A Amo, T C H Liew, C Adrados, E Giacobino, A V Kavokin, and A Bramati, {Phys. Rev. B {\bf 80}, 165325 (2009).}

\bibitem{wal_review_SST}M.M. Glazov and L.E. Golub, Semicond. Sci. Technol. \textbf{24}, 064007 (2009).

\bibitem{polarit_prb}M.M. Glazov and L.E. Golub, Phys. Rev. B \textbf{77}, 165341 (2008).


\bibitem[{\citenamefont{Gurioli et~al.}(2005)\citenamefont{Gurioli, Bogani,
  Cavigli, Gibbs, Khitrova, and Wiersma}}]{gurioli:183901}
\bibinfo{author}{\bibfnamefont{M.}~\bibnamefont{Gurioli}},
  \bibinfo{author}{\bibfnamefont{F.}~\bibnamefont{Bogani}},
  \bibinfo{author}{\bibfnamefont{L.}~\bibnamefont{Cavigli}},
  \bibinfo{author}{\bibfnamefont{H.}~\bibnamefont{Gibbs}},
  \bibinfo{author}{\bibfnamefont{G.}~\bibnamefont{Khitrova}}, \bibnamefont{and}
  \bibinfo{author}{\bibfnamefont{D.~S.} \bibnamefont{Wiersma}},
  \bibinfo{journal}{Phys. Rev. Lett.} \textbf{\bibinfo{volume}{94}},
  \bibinfo{eid}{183901} (\bibinfo{year}{2005}).
  
\bibitem{liew} T. C. H. Liew, C. Leyder, A. V. Kavokin, A. Amo, J. Lefr\`{e}re, E. Giacobino, and A. Bramati, Phys. Rev. B {\bf 79}, 125314 (2009).

\end{thebibliography}
\end{document}